\documentclass{an}
\usepackage{graphicx}
\usepackage{times}
\begin{document}

\Pagespan{1}{}
\Yearpublication{Year of publication}
\Yearsubmission{Year of submission}
\Month{Month of publication}
\Volume{Volume}
\Issue{Issue}
\DOI{DOI}
\title{Induced MHD oscillations of fine loop structures located in coronal hole.}
\author{S. Anfinogentov \and R. Sych \and D. Prosovetsky }
\institute{Institute of Solar-Terrestrial Physics SB RAS Irkutsk, Russia}
\abstract{Many phenomena seen in solar atmosphere are connected with underphotospheric processes.   MHD oscillation and other dynamical processes observed in quit sun objects are, for example, these phenomena. In our research we use Stereo Behind EUV(171\AA \ and 304\AA) observations to study dynamical processes in quiet Sun  regions. We found an eruption event in a coronal bright point on the border of a small coronal hole close to the center of the Solar disk. Several oscillating loops  became visible  inside of the  hole after the eruption. We suppose that these loops oscillation were induced by the eruption process in coronal bright point. The nature of interaction transfer  agent is unclear because any propagating disturbance was not detected. For data processing we used Pixelize wavelet filtration (PWF method) and Time-distance plots.  We measured the supposed interaction transfer speed and found it  about 2-3 km/s. We tried to find out the nature of the interaction transfer agent.}
\maketitle
\section{Introduction}
Dynamical events observed in quiet Solar corona are often connected with photospheric and underphotospheric processes. Innes et al.[\cite{Innes2009}] studied small scale eruptive events called Mini CME and showed that they are connected with supergranular flows. 

For our research  we  use data obtained by STEREO mission. It's on-board instrument SECCHI provides the combination of high spatial resolution  good sensibility and relatively high cadence. This allows us to study dynamical processes in small scale quiet Sun objects, such as coronal bright points and small coronal loops. 
\section {Observations}
\begin{figure*}
\includegraphics[width=130mm,height=80mm]{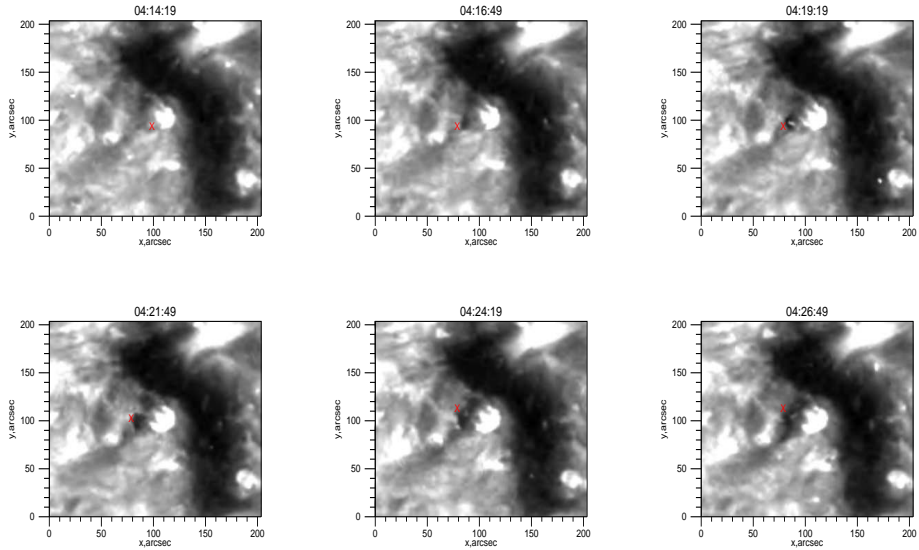}
\caption{6 subsequent EUV 171\AA \  images showing one of the observed eruption processes. The "X" sign shows propagating coronal dimming visible near the coronal point}
\label{fig_eruption}
\end{figure*}

\begin{figure}
\includegraphics[width=80mm,height=28mm]{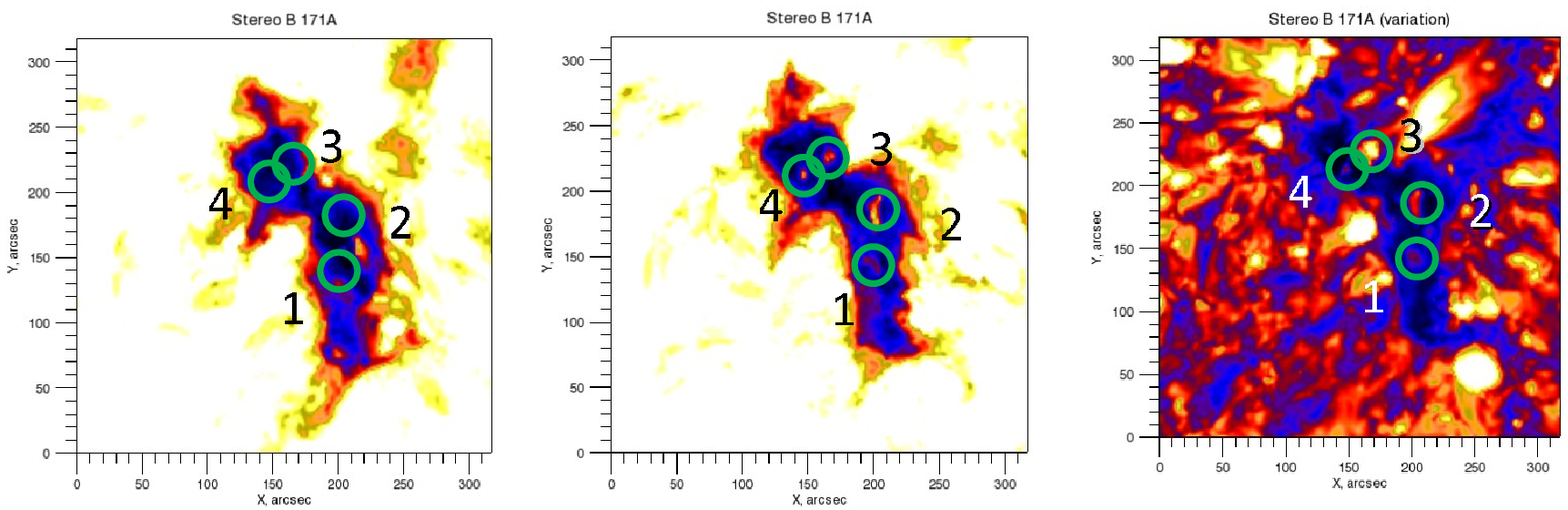}
\caption{EUV images of the hole. Left: 02:00 UT (before eruption, oscillating loops are not visible); middle: 07:52 UT (after eruption, oscillating loops have become visible); right: Variation map (02-12 UT) shows hot places where dynamical processes are present)}
\label{fig_loops}
\end{figure}

We use STEREO  171\AA \ image sequence from 02:00 till 12:00   July 7, 2010.   The 171\AA  \  observations have the highest cadence (2.5 min) for this date. 
\begin{figure}
\includegraphics[width=80mm,height=40mm]{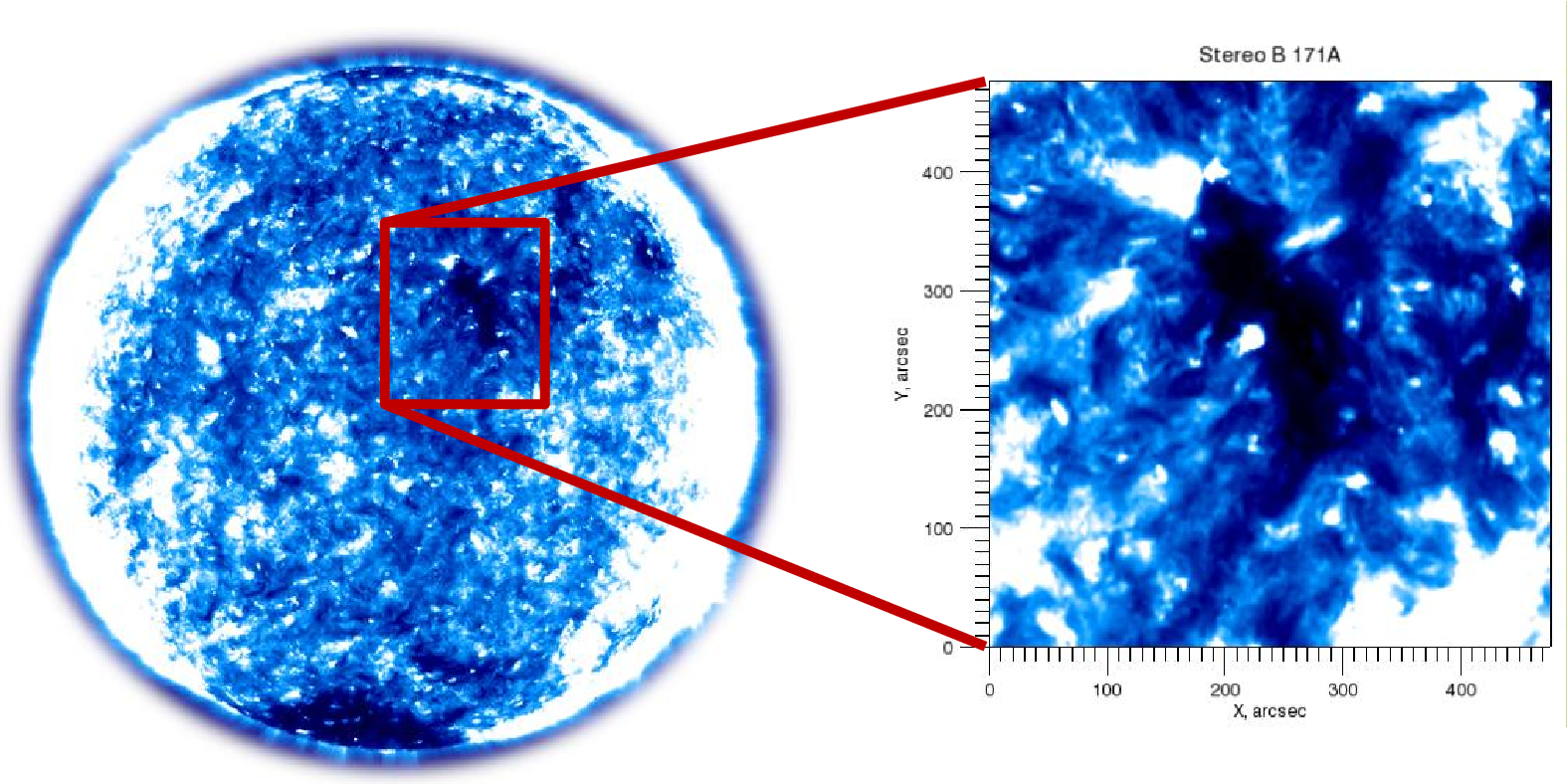}
\caption{STEREO Behind EUV (171\AA) image of the sun. Red square shows  the object to analyze (coronal hall)}
\label{fig_sun}
\end{figure}
We select a small coronal hole close to the center of the Solar disk on STEREO Behind images(fig \ref{fig_sun}).  A coronal bright point is located on the left border of the hole. We detected a sequence of eruptive events in the coronal point at 04:16, 05:51 and 06:54 UT. And at last the coronal bright point disappeared from EUVI images (08:36 - 08:44 UT). These events are illustrated on fig. \ref{fig_eruption}.
 Four oscillating loops appeared inside of the coronal hole after eruption events. The images of the hole before and after eruption events are given  on fig. \ref{fig_loops}.

\begin{figure*}
\includegraphics[width=160mm,height=32mm]{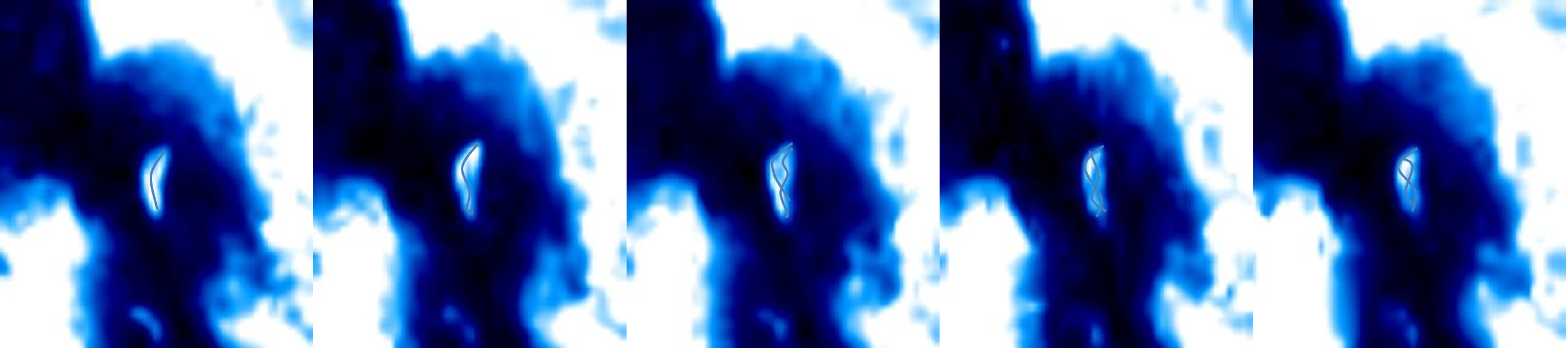}
\caption{5 subsequent EUV images illustrating oscillations in the small coronal loop number 2. 
The supposed fine structure of the loop is showed with solid lines.
}
\label{fig_loop2}
\end{figure*}

\begin{figure}
\includegraphics[width=80mm,height=50mm]{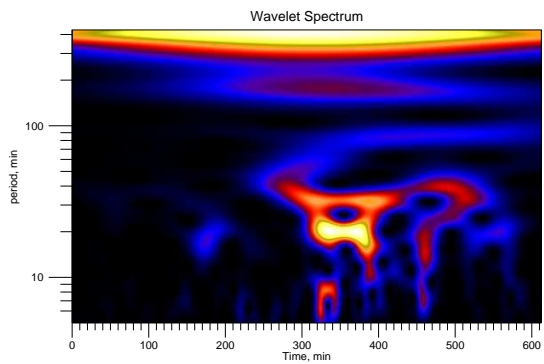}
\caption{Wavelet power spectrum of the EUV emission flux from oscillating loop 2.
}
\label{fig_wavelet}
\end{figure}

We found 2 types of oscillations inside the loops. The first type is propagating along the loop wave which is visible as brightness variation. The second type can be a kink mode. Fig. \ref{fig_loop2} illustrates  oscillations in the loop number 2.
For each loop we have measured oscillation periods and starting time by the wavelet transform technique. You can find detailed information about wavelet transform and its application for studying oscillation in solar atmosphere  in paper [\cite{Sych2007}]. The wavelet spectrum of the  emission from loop 2 is shown on Fig. \ref{fig_wavelet}.
 We can see two horizontal stripes on the spectrum. They correspond to oscillation with periods of 20 and 30 min. We also can determine when oscillation started and when they stopped. Oscillations start and stop time as well as their periods are presented in the table \ref{tab_osc}.
\begin{table}
\caption{The oscillation parameters for all visible loops inside the coronal hole.}
\label{tab_osc}
\begin{tabular}{ccccc}\hline
Loop number & 1 & 2 &3 & 4 \\
\hline
Start time, UT & 06:10 & 06:10 & 08:00 & 06:30 \\
Periods, min & 30 & 20 and 35 & 25 and 60 & 15 and 35 \\
\hline
\end{tabular}
\end{table}

\section {Interaction transfer speed measurement}
\begin{figure}
\includegraphics[width=80mm,height=30mm]{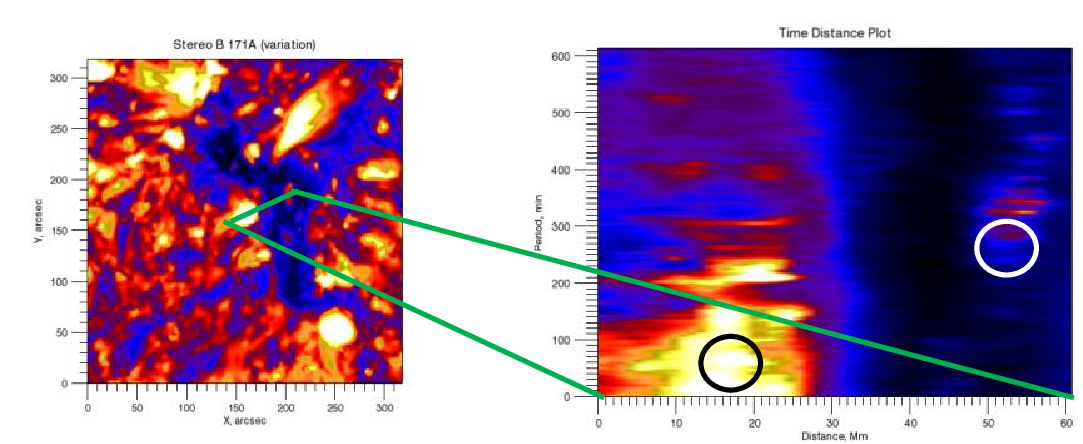}
\caption{Time shifts and distance measurements. Left: Variation map of the observational data; green line shows the profile direction. Right: time-distance plot through the line connecting eruption location and oscillating loop.  Black circle  shows  eruptive event location, white circle shows the start of the loop oscillation.}
\label{fig_timedist}
\end{figure}

An eruptive event was detected on the border of the coronal hole. It was followed by the loop oscillations under discussion. We assume that the eruption caused loop oscillation inside coronal hole. We measured distances  between eruption center and oscillating loops for investigating this hypothesis. We also measured time shifts between the  eruption start and the start of loops oscillations. Distances and time shifts allow us to calculate interaction transfer speeds. 
We use "time – distance plots" to measure distances and time shifts.
This method is illustrated on Fig. \ref{fig_timedist}.
 We use the intensity profile of the data cube along the line connecting the coronal bright point and the oscillating loop. Horizontal axis is distance, vertical axis is time. Using  the time-distance plot, we can measure distances and time shifts between processes. Measured time shifts, distances and interaction transfer speeds are given in the table \ref{tab_speed}.  

\begin{table}
\caption{Measured time shifts, distances and interaction transfer speeds for four oscillating loops visible inside of the coronal hole.}
\label{tab_speed}
\begin{tabular}{ccccc}\hline
Loop number & 1 & 2 &3 & 4 \\
\hline
Time shift, min & 200 & 200 & 300 & 250 \\
Distance, Mm & 32 & 35 & 40 & 34 \\
Interaction speed,$km\cdot s^{-1}$ & 2.7 & 2.9 & 2.2 & 2.3 \\
\hline
\end{tabular}
\end{table}

\section {Discussion}
Longitude and kink oscillation have been detected in small  scale loop structure inside coronal hole. The different loops have different oscillation periods  about 15-30 minutes.  It is shown that these oscillations  appear in several places of coronal hole after eruptive process in coronal bright point on the border of the hole. We also note that loop structures had not been visible on EUV images before oscillations started. We have measured distances between Eruptive event and oscillating loops. We also have measured time shifts between oscillation start in each loop structures and eruptive event in coronal bright point.
 The speed of supposed propagating agent was found to be almost the same for all loops (2-3 $km\cdot s^{-1}$). We made a conclusion that the oscillations were initiated by eruptive process in coronal bright point and the interaction transfer agent is same for all four loops. Interaction transfer speed  is much lower than estimated sound speed in solar corona (173 $km\cdot s^{-1}$ for 171\AA \ line forming temperature). It means that only slow mode  MHD waves are allowed. But this mode can't propagate  across magnetic field lines (in coronal hole magnetic field lines are open and almost perpendicular to the solar surface and to the direction of the supposed disturbance propagation). We have not detected any propagating atmospheric disturbance in 171\AA \ , 195\AA \  and 304\AA \ data. This is an argument for the underphotospheric or photospheric nature of the disturbance agent. We made a conclusion that the transfer agent can be photospheric gravitational wave or underphotospheric propagating  disturbance . 
\newpage

\bibliography{quiet}
\end{document}